\documentclass[prl,aps,amssymb,showpacs,twocolumn]{revtex4}
\usepackage{amsmath}
\usepackage{amssymb}
\usepackage{amsthm}
\usepackage{amsfonts}
\usepackage{listings}
\usepackage{enumerate}
\usepackage{latexsym}
\usepackage[dvips]{graphicx}
\usepackage{psfrag}
\usepackage{bm}
\usepackage{graphicx}
\usepackage{subfigure}

\newcommand{\beq}{\begin{equation}}
\newcommand{\eneq}{\end{equation}}

\newcommand{\bra}[1]{\left\langle#1\right|}
\newcommand{\ket}[1]{\left|#1\right\rangle}

\input{epsf}

\begin{document}

\tolerance 10000

\newcommand{\vk}{{\bf k}}


\title{The entanglement gap and a new principle of adiabatic continuity}

\author{R. Thomale$^{1}$, A. Sterdyniak$^{2}$, N. Regnault$^{2}$, and
  B.  Andrei Bernevig$^{1}$} 
\affiliation{$^1$ Department of Physics,
  Princeton University, Princeton, NJ 08544} 
\affiliation{$^2$ Laboratoire Pierre Aigrain, ENS and CNRS, 24 rue Lhomond, 75005 Paris, France}

\begin{abstract}
  We give a complete definition of the entanglement gap separating
  low-energy, topological levels, from high-energy, generic ones, in
  the "entanglement spectrum" of Fractional Quantum Hall (FQH) states.
  By removing the magnetic length inherent in the FQH problem - a
  procedure which we call taking the "conformal limit", we find that
  the entanglement spectrum of an incompressible ground-state of a
  generic (i.e. Coulomb) lowest Landau Level Hamiltonian re-arranges
  into a low-(entanglement) energy part separated by a \emph{full} gap
  from the high energy entanglement levels. As previously observed
  \cite{li-08prl010504}, the counting of these levels starts off as
  the counting of modes of the edge theory of the FQH state, but
  quickly develops finite-size effects which we show can also serve as
  a fingerprint of the FQH state. As the sphere manifold where the FQH
  resides grows, the level spacing of the states at the same angular
  momentum goes to zero, suggestive of the presence of
  \emph{relativistic} gapless edge-states. By using the adiabatic
  continuity of the low entanglement energy levels, we investigate
  whether two states are topologically connected.
\end{abstract}

\date{\today}


\maketitle

Topological phases of matter generally lack local order parameters
that can distinguish them from trivial ones. Moreover, extracting the
topological order directly from the ground-state wavefunction is a
nontrivial task.  For incompressible states, several non-local
indicators of the topological nature of the phase, such as
ground-state degeneracy on compact high genus manifolds, the structure
of edge modes and their scaling exponents, as well as quantum
dimension analysis exist, but still do not fully describe the
topological phase. The measure of choice has so far been the
entanglement entropy (EE), especially its topological
part~\cite{Kitaev-06prl110404,Levin-05prb045110}. For a given state
$\ket{\Psi_0}$ and according density matrix $\rho = \ket{\Psi_0}
\bra{\Psi_0}$, let the Hilbert space be decomposed as a direct product
$\mathcal{H}=\mathcal{H}_A \otimes \mathcal{H}_B$. Defining $\rho_A
\equiv \text{Tr}_B [\rho]$, the EE with respect to the partitioning
$(A,B)$ is defined by $S_A=-\text{Tr}_A [\rho_A \text{ln} \rho_A]$.
For two-dimensional quantum systems, except in special cases where
analytical solutions can be found~\cite{stephan-09cm}, extracting the
topological part of the EE becomes a highly nontrivial (and almost
impossible) task.

While the EE is just one number, it was recently proposed and
numerically substantiated~\cite{li-08prl010504} that the entanglement
spectrum (ES), i.e. the full set of eigenvalues of $\rho_A$,
understood as a geometric partition of the quantum Hall
sphere~\cite{haldane83prl605}, is a better indicator of topological
order in the ground state of FQH systems. Writing the eigenvalues as
the spectrum of a fictitious Hamiltonian, $\rho_A=\text{exp}(-H)$,
where one can think of the $H$ eigenvalues $\xi$ as a quasi-energy (or
entanglement energy), Li and Haldane~\cite{li-08prl010504} showed that
the low quasi-energy spectrum for generic gapped $\nu = 5/2$ states
exhibits a universal structure, related to conformal field theory. A
few of the eigenvalues displaying this CFT counting are separated from
a non-universal high energy spectrum by an entanglement gap which was
conjectured to be finite in the thermodynamic (TD) limit
~\cite{li-08prl010504}. This gap itself was proposed as a
"fingerprint" of the topological order present. It was subsequently
shown that the ES can meaningfully distinguish among states which have
similar finite size overlap with each other, but different edge
structures~\cite{regnault-09prl016801}.  Recently, the ES was found to
detect topological order in gapless spin chains~\cite{thomale-09es}.
\begin{figure*}[t]
  \begin{minipage}[l]{0.41\linewidth}
\hspace{-5pt}
    \psfrag{B}{$\xi$}
    \psfrag{A}{$L_{z}^{A}$}
    \includegraphics[width=\linewidth]{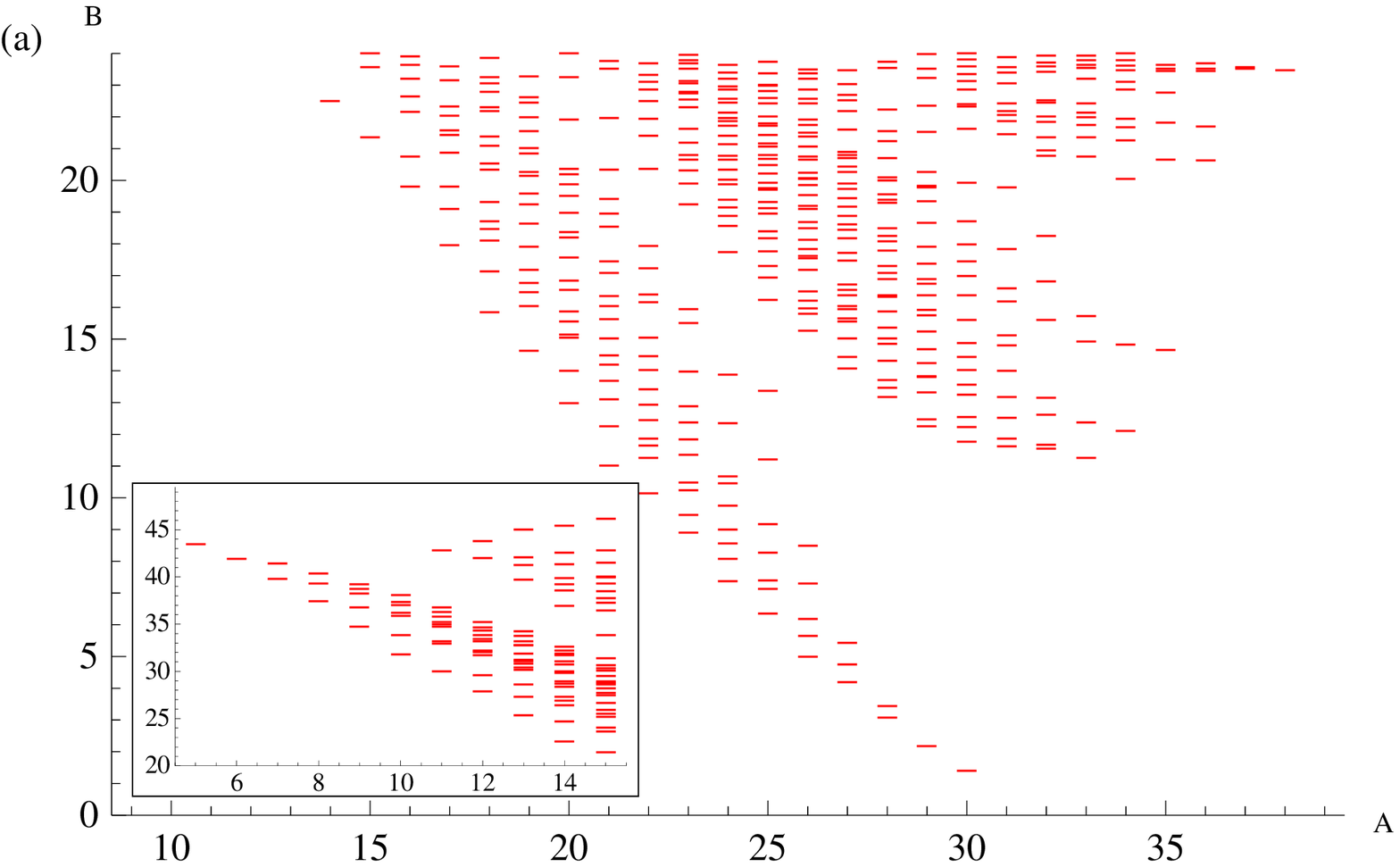}
  \end{minipage}
\hspace{10pt}
  \begin{minipage}[l]{0.41\linewidth}
\hspace{5pt}
    \psfrag{B}{$\xi$}
    \psfrag{A}{$L_{z}^{A}$}
    \includegraphics[width=\linewidth]{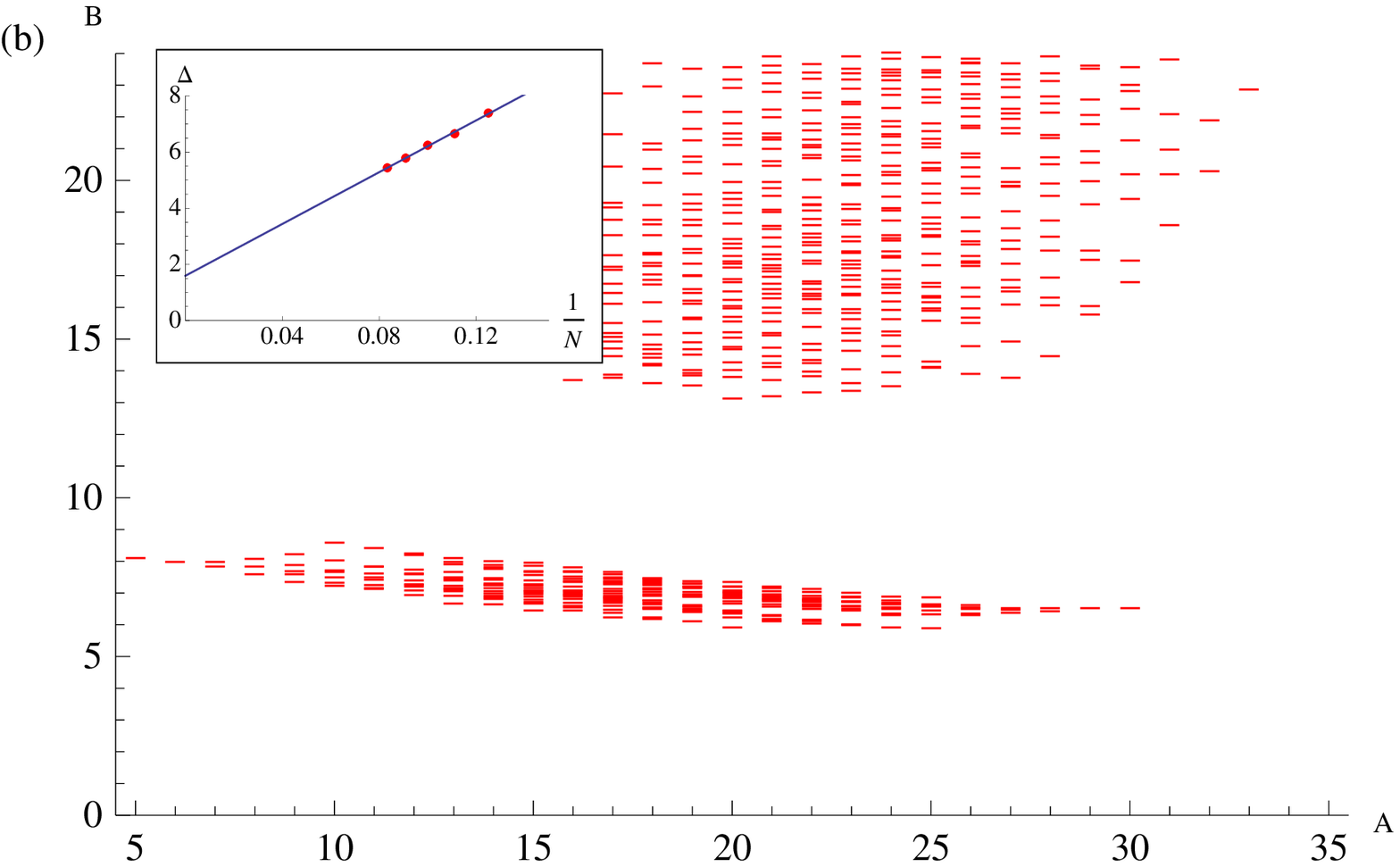}
  \end{minipage}
  \caption{Entanglement spectrum for the $N=11$ bosons, $N_\phi=20$,
    $\nu=1/2$ Coulomb state on the sphere. The cut is such that
    $l_A=10$ orbitals and $N_A=5$ bosons.  (a) Standard normalization
    on the quantum hall sphere. The inset show the remainder part of
    the spectrum where the entanglement levels exceed $\xi=24$. (b)
    CL normalization. We observe that the CL
    separates a set of universal low-lying energy states,
    which allows an unambiguous definition of the entanglement gap
    over all $L^A_z$ subsectors as
    the minimal difference between the highest energy CFT state and
    lowest generic state. The inset in (b)
    shows the the finite size scaling of the entanglement gap for the
    Coulomb state, which remains finite in the TD limit.}
\label{coubos-ba}
\vspace{0pt}
\end{figure*}

An unambiguous definition of the entanglement gap is still an open
question. As the angular momentum of the northern hemisphere grows,
i.e. $L_z^A$, the $z$ component of the angular momentum of the
complementary region A, is reduced, the entanglement gap collapses: in
finite sizes, and for good incompressible states with large gap this
happens at roughly $4-5$ momenta below that of the minimum angular
momentum for the hemisphere where the cut was made (see Fig.
[\ref{coubos-ba}a]).  For these $4-5$ momenta, the state shows the
counting of the edge modes of its corresponding
CFT~\cite{li-08prl010504}, and deviates from this counting once the
entanglement spectrum "feels" the edge (north pole) of the system.
For other FQH states (such as the $\nu=5/2$ Pfaffian state), the
entanglement gap as currently defined is rather small and already
disappears after $2-3$ angular momenta above the minimum one. As we
raise the angular momentum of the northern hemisphere, the ES levels
form a continuum of states, which previously led to the conclusion
that these levels are not useful for determining the character of a
FQH state. Also, if we assume the conjectured mapping of entanglement
energies to edge mode energies, it is unclear why states at the same
angular momentum would have different entanglement energies, as the
dispersion on the edge is relativistic.

In this paper, we give a precise definition of the entanglement gap.
We notice that the previous applications of the
ES~\cite{li-08prl010504,regnault-09prl016801} contained the geometry
of the Landau orbitals on the manifold in question (sphere), and hence
implicitly had involved the magnetic length. Inspired by our previous findings on spin chains~\cite{thomale-09es}, by removing the magnetic
length from the problem, we obtain the "conformal limit" (CL) of the FQH
polynomial. For model FQH states, the CL has the
desirable property that the spacing between entanglement eigenvalues
at the same angular momentum goes to zero very quickly as the sphere
is enlarged, thus cementing the relation between entanglement energies
and edge mode energies. The low-lying levels start by showing the
universal CFT counting but then exhibit finite size effects. For
generic FQH states, obtained by diagonalizing the
Coulomb Hamiltonian, the entanglement spectrum in the CL
exhibits a \emph{full} gap between \emph{all} the model levels and the
generic, high-energy Coulomb ones. This shows that not only the
CFT-like levels are important in the determination of a state: the
levels which exhibit finite-size effects are also  a
fingerprint of the state.

Diagonalizing a many-body Hamiltonian invariably introduces normalization
factors of the non-interacting many-body states which depend on the
specific geometry of the underlying manifold. In particular, these
factors contain the information about the extent of the
Landau orbitals in space, and depend on the magnetic length of the
problem. Stated differently, this type of normalization relies on the
curvature, i.e. a local quantity of the manifold. By contrast, the
CL should by definition contain no real length-scale. We
are led to the conclusion that the best way to analyze a FQH
polynomial obtained from the diagonalization of any Hamiltonian is to
\emph{un-normalize} it and strip it down of its magnetic length
information. 
We now exemplify this procedure for the sphere geometry.  Free boson
states are spanned by the monomials $m_{\lambda} =\frac{1}{\prod_{j}
  n_j!}  \text{Per}(z_i^{\lambda_j})$, where $i$ runs over the number
of particles $N$ and $j$ over the number of orbitals, and $n_j$
denotes the multiplicity of occupation of the $j$th orbital. $\lambda$
defines a partition of the angular momentum $\lambda_j$ of different
occupied orbitals, and $\text{Per}$ denotes the permanent state with
single particle positions $z_i$. The $m_{\lambda}$ are free
many-particle states that are \emph{unnormalized}.
\begin{figure}[t]
  \begin{minipage}[l]{0.48\linewidth}
    \psfrag{B}{$\xi$}
    \psfrag{A}{$L_{z}^{A}$}
    \includegraphics[width=\linewidth]{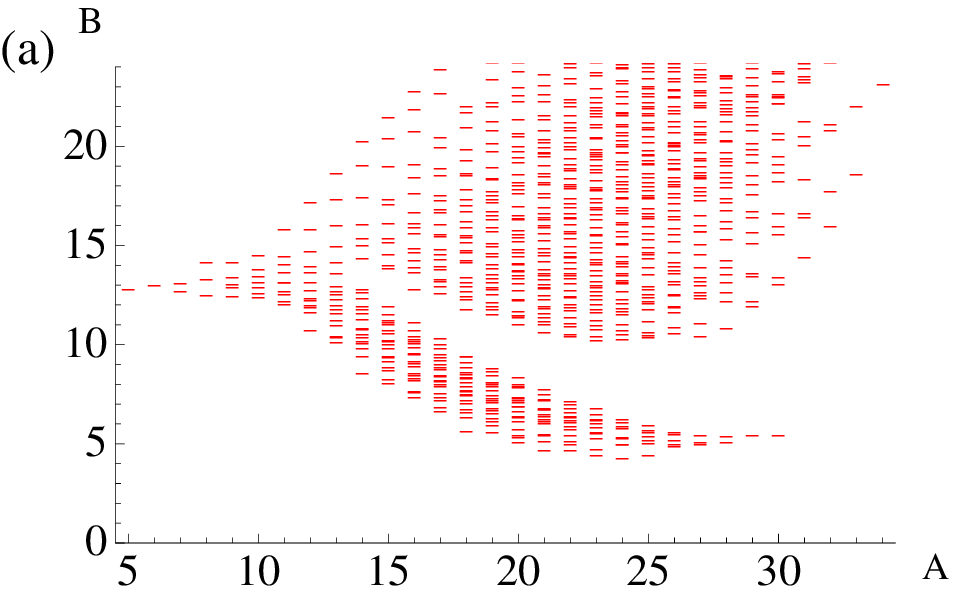}
  \end{minipage}
\hspace{2pt}
  \begin{minipage}[l]{0.48\linewidth}
    \psfrag{B}{$\xi$}
    \psfrag{A}{$L_{z}^{A}$}
    \includegraphics[width=\linewidth]{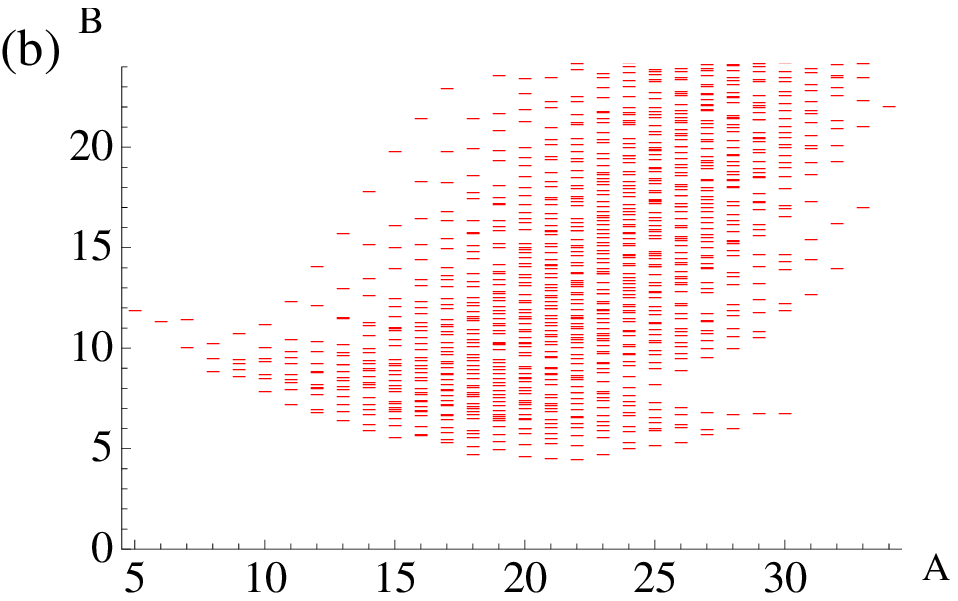}
  \end{minipage}
  \caption{ Entanglement spectrum at filling $\nu=1/2$ for the $N=11$
    bosons and the ground state of Coulomb interaction with a modified
    short range component by some $\delta V_0$ pseudopotential.  Here
    $N_\phi=20$, $N_A=5$ and $l_A=10$. Left panel (a) is obtained for
    $\delta V_0=-0.35$ where the gap starts closing. Right panel (b)
    is for $\delta V_0=-0.425$, close to transition to an compressible
    $L \ne 0$ state.}
\label{comp}
\vspace{-0pt}
\end{figure}
When one diagonalizes a many-body Hamiltonian, the expansion of the
interacting wavefunction is in \emph{normalized} free many-body states
$\cal{M}_\lambda$, which differ from the unnormalized basis above
through normalization factors that contain information
about the geometry of the manifold and the magnetic length.
\begin{figure*}[t]
  \begin{minipage}[l]{0.23\linewidth}
    \psfrag{B}{{\small$\xi$}}
    \psfrag{A}{{\small$L_{z}^A$}}
    \includegraphics[width=\linewidth]{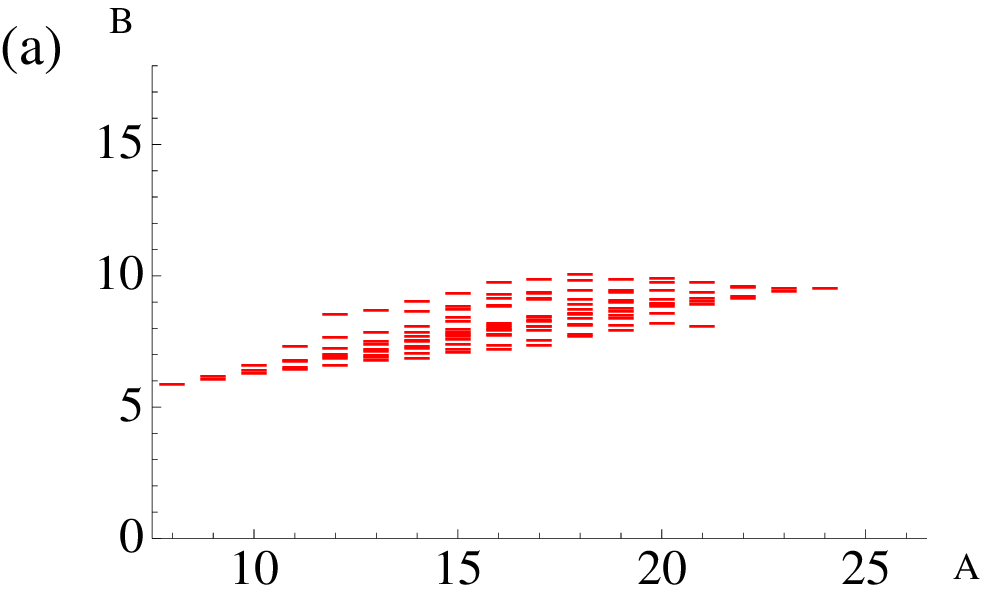}
  \end{minipage}
\hspace{5pt}
  \begin{minipage}[l]{0.23\linewidth}
    \psfrag{B}{{\small$\xi$}}
    \psfrag{A}{{\small$L_{z}^A$}}
    \includegraphics[width=\linewidth]{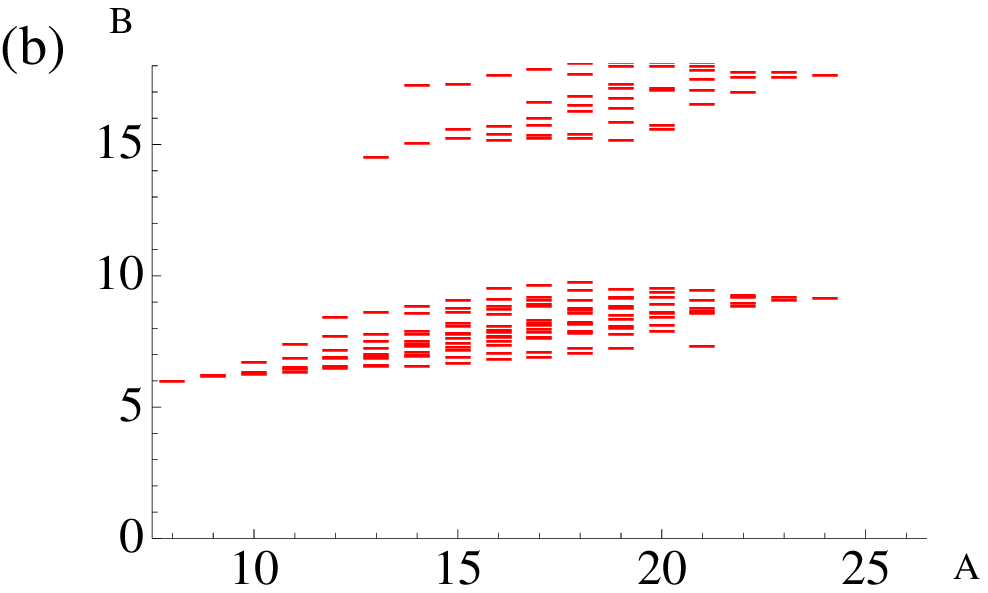}
  \end{minipage}
\hspace{5pt}
  \begin{minipage}[l]{0.23\linewidth}
    \psfrag{B}{{\small$\xi$}}
    \psfrag{A}{{\small$L_{z}^A$}}
    \includegraphics[width=\linewidth]{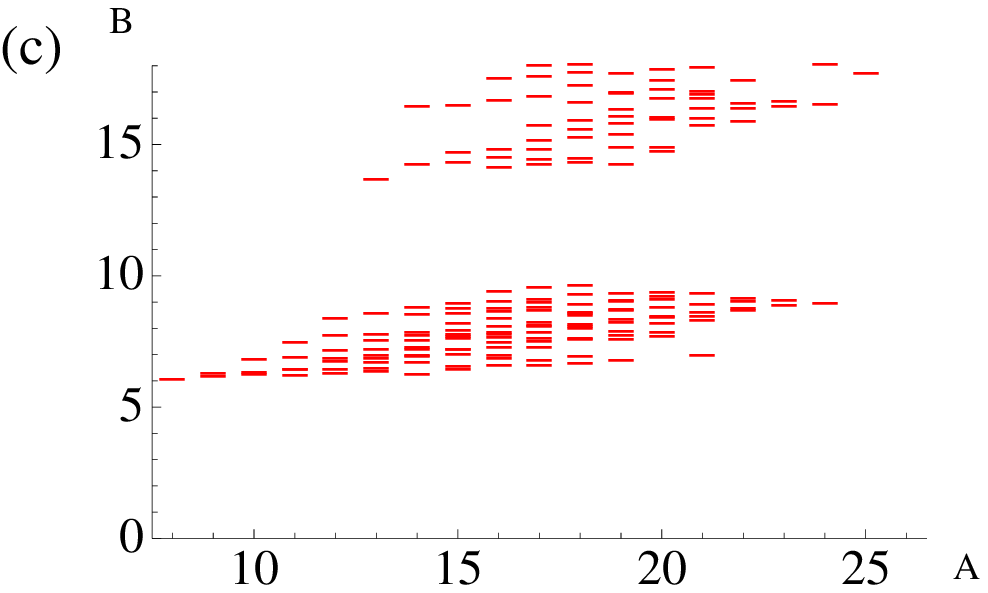}
  \end{minipage}
\hspace{5pt}
  \begin{minipage}[l]{0.23\linewidth}
    \psfrag{B}{{\small$\xi$}}
    \psfrag{A}{{\small$L_{z}^A$}}
    \includegraphics[width=\linewidth]{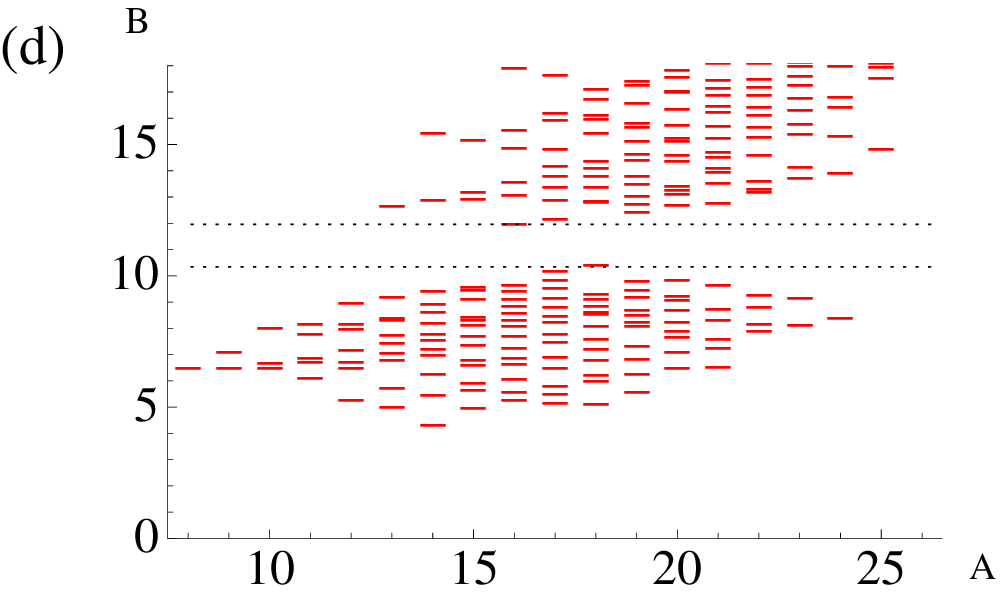}
  \end{minipage}
  \caption{Entanglement spectra for the $\nu=1$ ground state
    of~\eqref{pfaff} for $\lambda=0$ (a, pure MR state), $0.15$ (b),
    $0.5$(c), and $1$ (d, pure delta ground state) for $N=14$ bosons,
    $N_\phi=12$, and the cut specified by $N_A=7$, $l_A=6$.  We
    observe that the entanglement gap shrinks, but retains a finite
    value up to the pure two-body potential. The dashed lines refer to
    sector $L_z^A=17$ where the minimum difference between the highest
    CFT level and lowest generic level, i.e. the entanglement gap, is
    found. The low energy spectrum retains the identical structure and
    is adiabatically connected.}
\label{pfaffsweep}
\vspace{-0pt}
\end{figure*}
On the sphere of radius $R$ the normalization of $m_\lambda$ is given
by~\cite{fano-86prb2670}
\begin{equation}
{\cal{N}_{\lambda}}^{\text{sphere}}=\left(  \frac{4 \pi}{( 2S +1)!}
\right)^N \frac{N!}{\prod_{j=0}^{\lambda_1} n_j!} \prod_{i=1}^N
\lambda_i! (2 S - \lambda_i)!,
\label{sphere}
\end{equation}
\noindent where $n_j$ is the multiplicity of the $j$th orbital in the
decreasingly ordered partition $\lambda = (\lambda_1, \lambda_2 ...
\lambda_N)$, where $\lambda_i \in [0, 2 S] $ is the angular momentum
of the Landau orbitals . We set the partition to be padded, such that,
if the initial partition has $l_\lambda$ number of elements non-zero
then all the rest $\lambda_{l_\lambda +1}...\lambda_N=0$. The number
of orbitals is conventionally given by $2S+1$, where $N_{\phi}=2S$ is
the magnetic flux.
We then apply the
transformation
\begin{equation}
{\cal{M}_{\lambda}}=  m_{\lambda}/\sqrt{ {\cal{N}_{\lambda}}^{\text{sphere}}} \label{eq:unsphere}
\end{equation}
\noindent to write the state as a function of the \emph{unormalized}
free boson many -body states.  As an example, in the new basis, the
\emph{unnormalized} Laughlin state for two particles reads $m_{(2,0)}
- 2 m_{(1,1)}$. In the unnormalized basis, all the coefficients of the
Laughlin state are integers. It is the basis which shows significant
structural information about the
polynomial~\cite{bernevig-08prl246802,bernevig-09cm0902}. For bosons,
the last step is to normalize each of the new free many-body
states by the square-root of the product of the factorials of their
bosonic multiplicities $m_{\lambda}=
\frac{1}{\sqrt{\prod_{j=0}^{\lambda_1} n_j!}}\tilde{m}_{\lambda}
\label{eq:unfac}$. Once expressed in this new basis (the conformal
limit), we calculate the ES for the ground-states of different
Hamiltonians.

As the first example, Fig.~\ref{coubos-ba} illustrates the conformal
limit transformation of the bosonic $\nu=1/2$ Coulomb state.  The
sphere is partitioned into two parts $A$ and $B$ in the orbital space
which mimics the geometrical partition~\cite{li-08prl010504}.  The
region $A$ is made of the $l_A$ first orbitals, starting from the
north pole. In region $A$, the total number of particles $N_A$ and the
projection of total angular momentum $L_z^A$ are good quantum numbers
to define the different sectors of the ES. We define the ES to be the
minimal difference between the highest energy CFT state and the lowest
energy generic state of all different sectors.
As shown in Fig.~\ref{coubos-ba}, the spectrum cleanly rearranges in a
low entanglement energy part and a high entanglement energy part,
separated from each other by a homogeneous entanglement gap, unlike in
the case of the sphere geometry where the entanglement gap can be
defined for only a few $L_z^A$ values.  Moreover, the state counting
in the low energy part of the Coulomb spectrum exactly matches with
the pure Laughlin spectrum for each $L_z^A$ sector.

We investigated the behavior of the gap when going through a phase
transition toward a compressible state.  A previous study for the ES
in the sphere normalization has been done for
$\nu=1/3$~\cite{zozulya-08prb347}.  In a similar way, using the
Haldane pseudopotentials decomposition of the Coulomb interaction, we
modify the pseudopotential associated to the short range component by
some amount $\delta V_0$ to drive the system into a compressible
state.  Starting from the Coulomb interaction at $\nu=1/2$ for $N=11$
bosons, the transition occurs at $\delta V_0 \simeq -0.45$.
Fig.~\ref{comp} shows two particular values where the gap starts
closing ($\delta V_0 \simeq -0.35$) and close to the transition point
($\delta V_0 \simeq -0.425$). The (square) overlap with
the Laughlin state stays rather high (resp. 0.9895 and 0.9288).  With
such overlaps, one would conclude that we are still in the same
quantum phase. Here the ES gives a more precise insight and tends to
show that the transition may occur for larger $\delta V_0$. Still,
there is no proof that as soon as the gap closes in one $L_z^A$
sector, all topological properties are lost.


Our CL basis enables us to study whether different states
are entanglement adiabatically continuable to each other. We
conjecture that two states are entanglement adiabatically continuable
if we can find a path to go from one state to the other without
collapsing the full entanglement gap. If so, we conjecture that the
states have identical topological structure. Let us illustrate this
property with the example of ultracold neutral bosons in a rapidly
rotating atomic trap. In this regime, FQH states are realized through
the two-body hardcore interaction (see e.g.~\cite{cooper-08ap539}). We
will focus on the filling $\nu=1$ where there is strong
evidence~\cite{regnault-06jpb89} that the system is described by the
Moore-Read (MR) state~\cite{Moore-91npb362}. We define a one parameter
Hamiltonian that linearly interpolates between the three-body hardcore
interaction for which the MR state is the exact zero energy
state~\cite{greiter-91prl3205}, and the two-body hardcore interaction
\begin{equation}
{\mathcal{H}}_\lambda=(1-\lambda) \sum_{i<j<k} \delta(r_i-r_j) \delta(r_j-r_k) + \lambda 
\sum_{i<j} \delta(r_i-r_j),
\label{pfaff}
\end{equation}
Fig.~\ref{pfaffsweep} shows spectra for several values of $\lambda$.
We find that the spectra of the pure three-body hardcore potential and
the two-body hardcore Hamiltonian are entanglement adiabatically
connected within the ES.  Finite size scaling for the individual
Hamiltonians also shows that the entanglement gap, though smaller for
the Pfaffian case at $\nu=1$ than for the previously studied Laughlin
at $\nu=1/2$, persists in the TD. Even though the overlaps between the
ground state at $\lambda=1$ and the MR state are lower (0.8858 for
$N=14$) than the ones we have previously mentioned in the $\nu=1/2$
case close to the phase transition, in this case there is a clear
entanglement gap. This example clearly shows that high overlap is not
a good indicator of a possible entanglement gap.

\begin{figure}[t]
  \begin{minipage}[l]{0.48\linewidth}
    \psfrag{B}{$\xi$}
    \psfrag{A}{$L_{z}^{A}$}
    \includegraphics[width=\linewidth]{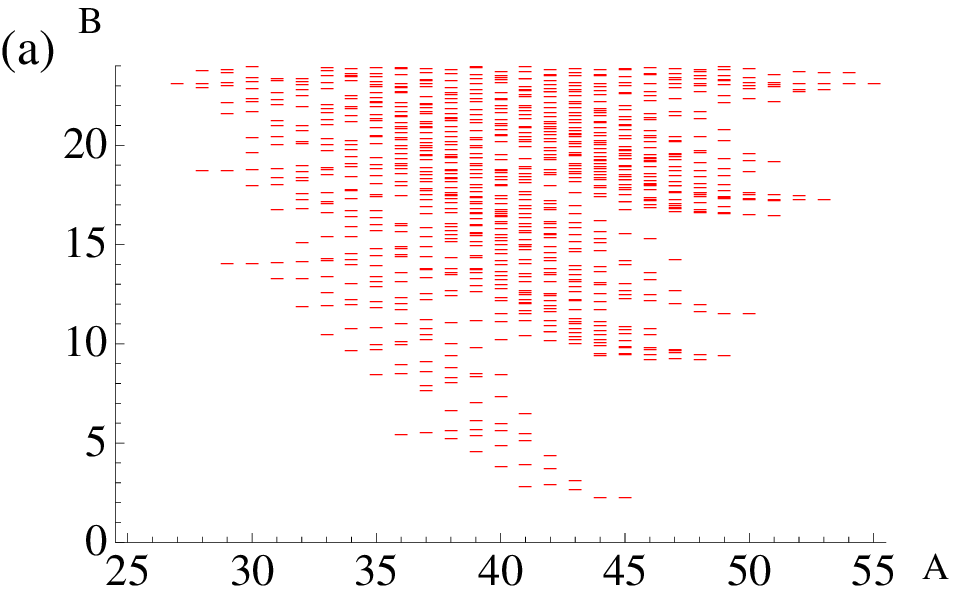}
  \end{minipage}
\hspace{2pt}
  \begin{minipage}[l]{0.48\linewidth}
    \psfrag{B}{$\xi$}
    \psfrag{A}{$L_{z}^{A}$}
    \includegraphics[width=\linewidth]{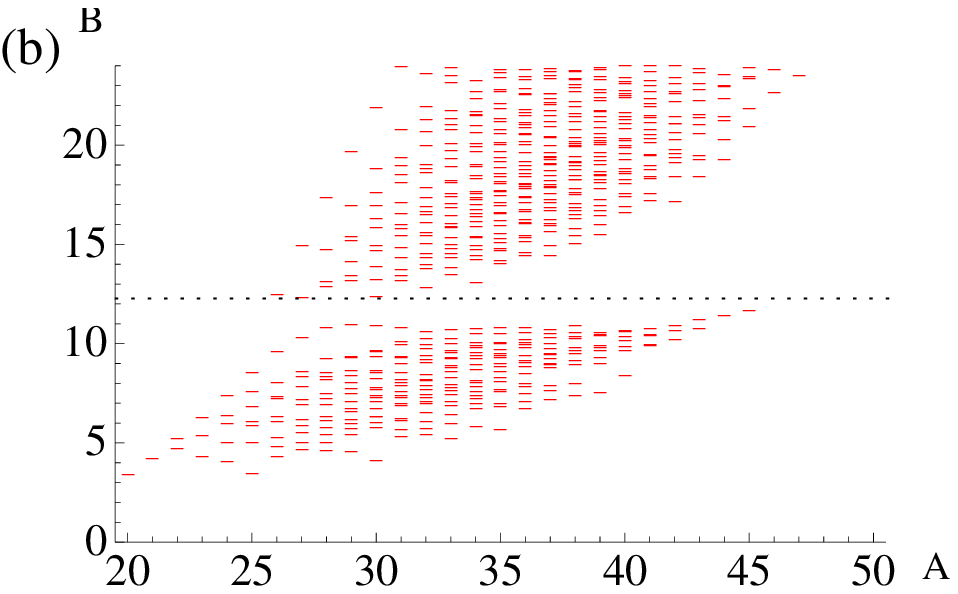}
  \end{minipage}
  \caption{Entanglement spectrum for $N=11$ fermions Coulomb state
    at $\nu=1/3$ filling, $N_\phi=30$, and $N_A=5$, $l_A=14$. As in
    the bosonic case shown in Fig.~\ref{coubos-ba},
    starting from the sphere normalization in (a), the CL
    basis in (b) clearly separates the universal states from the
    generic states in the spectrum indicated by the
    dashed line, and a clear entanglement gap over
    all $L^A_z$ subsectors emerges.}
\label{fermions}
\vspace{-0pt}
\end{figure}

For the case of the fermionic wavefunctions of a generic Hamiltonian,
one has to perform the same operations~\eqref{eq:unsphere}, while the
occupation multiplicity terms are trivial. In
Fig.~\ref{fermions}, we show the rearrangement of the ES of the
fermionic $\nu=1/3$ Laughlin state upon the CL basis
transformation. Notably, the
universal CFT level part associated with the pure Laughlin state
levels completely separates from the generic levels in the
CL.



\begin{figure}[t]
  \begin{minipage}[l]{0.99\linewidth}
    \psfrag{A}{\small$\langle\xi \rangle_m$}
    \psfrag{B}{\small$\frac{1}{N}$}
    \psfrag{C}{\small$\frac{1}{N}$}
    \psfrag{D}{\small$\Delta_3$}
    \includegraphics[width=\linewidth]{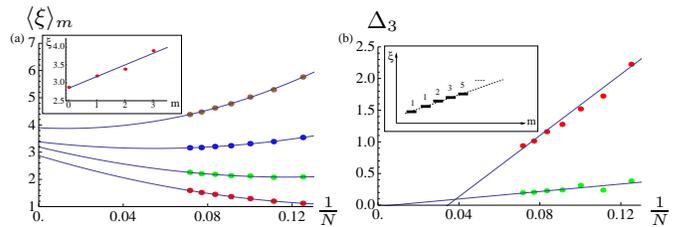}
  \end{minipage}
  \caption{(a) Fit of mean values $\langle\xi \rangle_m$ in the
    $\nu=1/2$ Laughlin state for different sectors
    $L^A_{z,\text{max}}-m$, for $m=0,1,2,3$. As system-size increases,
    the individual entanglement levels go up in "entanglement energy".
    The best fit is accomplished by an inverse quadratic fit to the
    number of bosons. The inset shows the extrapolated linear
    dispersion relation for the edge states.  (b) Scaling of the
    spread $\Delta_3$ of the $L^A_{z,\text{max}}-3$ sector for the
    sphere-normalized Laughlin state (red) and the Laughlin state in
    the CL (green). Both extrapolate to zero width in the
    TD limit. The inset shows the expected shape of the Laughlin ES
    spectrum for infinite system size - a linear dispersive set of
    states following the edge mode state counting and degenerate in
    the different $L_z^A$ sectors.}
\label{scalingslaughlin}
\vspace{-0pt}
\end{figure}

As first shown in ~\cite{li-08prl010504}, the counting of low energy
entanglement levels can be related to the CFT edge theory of the
state, which allows to identify topological bulk properties. We here
go further and investigate whether there is direct correspondence not
only between the counting of the levels but also between the actual
energies of the edge states and topological "entanglement energy"
levels.  From field theory~\cite{wen90prb12844}, edge states obey a
relativistic dispersion. If an entanglement level at
$L^A_z=L^A_{z,\text{max}}-m$ is identified to be related to an edge
state level, it obeys $E=\sum_{i} v (2\pi/L) k_i $, where $k_i$ is the
momentum of the individual field, $L$ the system length, $v$ the
velocity scale of the respective edge branch, and it holds $\sum_i k_i
= m$.
This implies that within one certain sector of total momentum $m$,
which in terms of entanglement levels would correspond to the sector
$L^A_{z,\text{max}}-m$, all entanglement levels corresponding to
different partitions of momentum $m$ on different edge fields should
have the same energy.
Thus, in the TD limit where all finite size effects are
absent, we conjecture that the spread of the universal low energy
entanglement states in each $L_{z}$ sector should shrink to zero, and
the overall slope from one momentum sector to the other obeys a linear
dispersion relation. We illustrate this for the $\nu=1/2$ Laughlin
state, for which the edge spectrum consists of one single bosonic
branch and where we can go to suitably large system sizes
(Fig.~\ref{scalingslaughlin}). We pick the highest $L_z^A$ sectors
$L_{z,\text{max}}^A-m$ up to $m=3$.

We first obtain the mean value of the low energy states in one
sector for the TD limit, and then extrapolate the dispersion relation
with respect to $m$. We find that the extrapolated dispersion is
linear within moderate error, confirming the relativistic behavior of
these entanglement levels. We also find that the spread of the low
energy levels shrinks to zero in the TD limit. While this holds in
geometry, the CL basis makes this feature become apparent
already for small system sizes, as the low energy levels in one sector
become significantly squeezed (Fig.~\ref{scalingslaughlin}).


We thank F.D.M. Haldane for insightful discussions. RT was supported
by a Feodor Lynen fellowship of the Humboldt
foundation. BAB is supported by an Alfred P. Sloan
Fellowship and by Princeton University.

{\it Note added.}  After this work has been submitted, we became aware
of a recent work by L\"auchli et al. on the defintion of entanglement
spectra for the Laughlin state on the torus
geometry~\cite{laeuchli-09tor}. It would be interesting to see what
our conformal limit gives when applied to their analysis.

\end{document}